

\input harvmac
\noblackbox
\def\lae{\raise-.5ex\vbox{\hbox{$\; <\;$}\vskip-2.9ex\hbox{$\; \sim\;$}}}
\def\gae{\raise-.5ex\vbox{\hbox{$\; >\;$}\vskip-2.9ex\hbox{$\; \sim\;$}}}
\def\slash#1{\raise.15ex\hbox{/}\kern-.57em #1}

\def\ie{{\it i.e.}}

\def\two#1{\raise1.35ex\hbox{$\leftrightarrow$}\kern-.88em#1}
\def\lefta#1{\raise1.35ex\hbox{$\leftarrow$}\kern-.61em#1}
\def\righta#1{\raise1.35ex\hbox{$\rightarrow$}\kern-.61em#1}
\def\Dslash{\raise.15ex\hbox{/}\kern-.77em D}
\def\np#1#2#3{Nucl. Phys. {\bf #1} (#2) #3}
\def\pl#1#2#3{Phys. Lett. {\bf #1} (#2) #3}

\def\pr#1#2#3{Phys. Rev. {\bf #1} (#2) #3}
\def\prd#1#2#3{Phys. Rev. D {\bf #1} (#2) #3}
\def\vsl{\raise.15ex\hbox{/}\kern-.57em v}
\def\ie{{\it i.e.}}

\def\Boxmark#1#2#3{\global\setbox0=\hbox{\lower#1em \vbox{\hrule height#2em
     \hbox{\vrule width#2em height#3em \kern#3em \vrule width#2em}%
     \hrule height#2em}}%
     \dimen0=#2em \advance\dimen0 by#2em \advance\dimen0 by#3em
     \wd0=\dimen0 \ht0=\dimen0 \dp0=0pt
     \mkern1.5mu \box0 \mkern1.5mu }

\Title{\vbox{\baselineskip12pt\hbox{BUHEP-93-14}\hbox{HUTP-93/A016}
\hbox{hep-ph/9306265}}}
{\vbox{\centerline{The Implications of Naturalness}
\vskip2pt\centerline{in Effective Field Theory}
\vskip2pt\centerline{on the Masses of Resonances}}}

\centerline{Michael J. Dugan$^{a,b,1}$}
\centerline{and}
\centerline{Mitchell Golden$^{a,2}$}
\footnote{}{$^a$Lyman Laboratory of Physics,
Harvard University, Cambridge, MA 02138}
\footnote{}{$^b$Boston University, Department of Physics, 590 Commonwealth
Avenue, Boston, MA 02215}
\footnote{}{$^1$dugan@physics.harvard.edu}
\footnote{}{$^2$golden@physics.harvard.edu}
\vskip .4in
\centerline{\bf ABSTRACT}

Many years ago Weinberg formulated a definition of ``naturalness'' for
effective theories: if an effective theory is to make sense, coefficients must
not change too much when the cutoff scale is changed by a factor of order 1.
As an example, we consider simple field theories in which an $O(N)$ symmetry
spontaneously breaks to $O(N-1)$.  We show that in these theories Weinberg's
criterion for a natural effective theory may be applied directly to the
$S$-matrix; it implies that the scale of new physics, beyond the Goldstone
bosons, may not be too large: there is always a particle or a cut of mass
below or about $4 \pi f / \sqrt{N}$.  We discuss the range of convergence of
the expansion of the chiral Lagrangian.  It appears to be impossible to
construct an underlying theory of the type considered here that fails to
satisfy Weinberg's criterion.

\Date{6/93}

\newsec{Introduction}

The origin of the hadron masses in QCD has always been something of a mystery.
While one can easily see why the pions are nearly massless --- they are
approximate Goldstone Bosons --- the precise value of the $\rho$ mass is
difficult to connect to the QCD lagrangian.  In recent papers\ref\us{R. S.
Chivukula, M. Dugan, and M.  Golden, \prd {47}{1993}{47}\semi R. S. Chivukula,
M. Dugan, and M. Golden, \pl{B292}{1992}{435}.}, it was shown that the upper
bound, $M$, for the mass of the lightest hadron in a QCD-like theory is reduced
as the number of light\foot{In this context the word ``light'' means small in
mass compared to $M$.} flavors $N_f$ is increased:
\eqn\natural{
M \approx {4 \pi f_\pi \over \sqrt{N_f}}~~,
}
where $f_\pi$ is the pion decay constant.  In theories in which the couplings
between the hadrons are weak (QCD in the limit of a large number $N_c$ of
colors is an example of such a theory) the masses of the hadrons may not come
close to saturating this bound.  However, the real world hadrons are strongly
coupled to each other and the bound \natural\ is nearly saturated.

The relationship \natural\ was derived by considering the chiral Lagrangian ---
an effective field theory of pions.  The scale at which the chiral expansion
breaks down is roughly $M$, if the low energy theory is ``natural''\ref\SS{M.
Soldate and R. Sundrum, \np {B340}{1990}{1}}, and the scale of breakdown of
chiral perturbation theory should be identified with a non-analytic structure
in the $S$ matrix, such as a resonance \us.  The notion of a natural effective
theory was first formulated by Weinberg \ref\weinnat{S. Weinberg, Physica {\bf
96A} (1979) 327\semi {\it see also} H.  Georgi and A. Manohar, \np {B234}
{1984}{189}.}.  An effective theory is natural if the sizes of the
coefficients are stable against $O(1)$ changes in the the size of the cutoff.
Equivalently, a coefficient in the effective lagrangian must not be small
compared to its $\beta$ function.

In this paper we will explore the significance of requiring that the low
energy reduction of a model have a natural effective theory (NET) in the sense
of Weinberg.  We will consider models in which the symmetry breaking pattern
is $O(N) \to O(N-1)$, rather than the $SU(N_f) \times SU(N_f) \to SU(N_f)$ of
QCD-like theories.  While such models are neither useful as models of hadrons
nor as symmetry breaking sectors for the electroweak theory\foot{other than for
$N=4$, which {\it is} the standard model!}, they do have the merit that some
of them become simple in the limit of large $N$.

By constructing toy models of this type, we show that the NET criterion
actually implies something about the $S$-matrix.  One need not even consider
the effective Lagrangian.  We will show that in a model in which there are an
arbitrary number of resonances in Goldstone boson scattering, the lightest
resonance is always lighter than $M$.  We then generalize to a model in which
the Goldstone boson scattering has a branch cut as well as a pole and we find
again that the mass of the new physics, the cut or the pole, will be less than
$M$.

In the models considered in this work, the underlying theories become sick if
the couplings are chosen so as to violate the bound.  Thus it appears to be
impossible to construct a theory that does not satisfy the NET criterion.
Whether this continues to be so in more complex, realistic theories remains an
open question.

The plan of this paper is as follows.  In section 2 we discuss the properties
of $O(N)$ models in general.  In section 3 we discuss their behavior in the
limit of large $N$.  We will consider a subclass of large-$N$ $O(N)$ models,
in which all of the physics appears in the $s$-channel of Goldstone boson
scattering.  In section 4 we discuss the domain of convergence of the
effective field theory and the relationship of this to the existence of poles
in scattering amplitudes.  In section 5 we discuss the meaning of the NET
criterion in such models.  Section 6 illustrates the NET criterion by applying
it to the various models mentioned above.  We then make some concluding
remarks in section 7.

\newsec{The O(N) Model}

A model in which the pattern of symmetry breaking is $O(N) \to O(N-1)$ will
possess $N-1$ Goldstone bosons, which form a vector representation of the
unbroken $O(N-1)$.  These particles (which we shall refer to as pions,
$\pi$) are the excitations of the degrees of freedom in the directions
corresponding to the different vacua of the $O(N)$ theory.  Consider elastic
scattering of two pions.  The most general form of the amplitude consistent
with Bose symmetry, crossing, and the unbroken $O(N-1)$ symmetry is
\eqn\general{
a(\pi^i \pi^j \to \pi^k \pi^l) =
  A(s,t,u) \delta^{ij}\delta^{kl}
+ A(t,s,u) \delta^{ik}\delta^{jl}
+ A(u,s,t) \delta^{il}\delta^{jk}~~,
}
where the function $A$ is symmetric in its last two arguments.

It is frequently desirable to express \general\ in terms of the three
invariant ``isospin'' channels of the unbroken $O(N-1)$.  Regarding the
initial state as a two index tensor in $i$ and $j$, we see that it may be
decomposed into its trace times the identity tensor, an antisymmetric tensor,
and a symmetric traceless tensor.  We will refer to these three invariant
amplitudes as isospin 0, 1, and 2 respectively, in analogy to the case of
$N=4$, the case of ordinary pions.  One obtains
\eqn\decomp{
\eqalign{
a_0 &= (N-1) A(s,t,u) + A(t,s,u) + A(u,s,t) \cr
a_1 &= A(t,s,u) - A(u,s,t) \cr
a_2 &= A(t,s,u) + A(u,s,t) ~~ .
}
}
Note that the scalar singlet channel is enhanced by a factor of $N$ while the
other two are not\foot{This is an amusing contrast to the more realistic case
of an $SU(N_f)_L \times SU(N_f)_R \to SU(N_f)_V$ pattern of symmetry breaking.
In that case, the analogue of $a_{11}$, the channel with the quantum numbers
of the $\rho$ meson, {\it is} enhanced by a factor of $N_f$.}.  These isospin
amplitudes may in turn be decomposed into partial waves
\eqn\wavedef{
a_{I\ell}(s) = {1\over 64 \pi} \int_{-1}^1 a_I(s, \cos\theta)
P_\ell(\cos\theta) d\cos\theta~~,
}
where $I=0,1,2$, and $P_\ell$ is the Legendre polynomial of order $\ell$.

The interactions of the pions at low energies may be described by the chiral
Lagrangian in the standard fashion \ref\CCWZ{S. Coleman, J. Wess, and B.
Zumino, \pr{177}{1969}{2239}\semi C. Callan, S. Coleman, J. Wess, and B.
Zumino, {\it ibid} {\bf 177} (1969) 2246.}.  The chiral Lagrangian is an
expansion in numbers of derivatives.  The lowest order term has two:
\eqn\Ltwo{
{\cal L}_2 = \half \partial^\mu \pi_i \partial_\mu \pi_i +
\half \partial^\mu (\sqrt{f^2-\pi^2}) \partial_\mu (\sqrt{f^2-\pi^2})~~.
}
Here $f$ is the decay constant of the pion.  This Lagrangian ${\cal L}_2$ is
non-linear and expanding the square root in powers of $\pi / f$ gives
vertices with arbitrary (even) numbers of pions.

Using this lowest order Lagrangian, we may calculate the low-energy behavior
of the scattering amplitudes of pions.  Since ${\cal L}_2$ is unique, the
result is independent of the exact details of the full theory.  One finds
\eqn\lowest{
A(s,t,u) = {s \over f^2}~~,
}
and therefore,
\eqn\lowamps{
\eqalign{
a_{00} =& {(N-2) s \over 32 \pi f^2} \cr
a_{11} =& {s \over 96 \pi f^2} \cr
a_{20} =& -{s \over 32 \pi f^2} ~~ .
}
}

To go beyond the threshold behavior of the pions --- to higher order in $s$,
$t$, and $u$ in the function $A$ above --- one must go beyond the tree-level
computations using ${\cal L}_2$ and include the calculation of loop diagrams.
Since the Lagrangian above is non-renormalizable, there will be counterterms
with more than two derivatives.  Accordingly, to go up in energy one must also
include the higher order terms in the chiral Lagrangian.  Unfortunately, the
coefficients of these terms in the Lagrangian depend in general on the physics
of the full theory and are not universal.

Since different full theories will have different high energy behavior, it is
not possible to say very much about the spectrum at high energies.
Nonetheless, as we shall show in the following sections, the NET criterion
does place some constraints on the physics, independent of particular
theories.

\newsec{The Limit of Large $N$}

An interesting perspective on the problem is obtained if one contemplates
theories in which there exists a good large-$N$ limit.  By this we mean that
it is possible to let $N \to \infty$ while keeping finite the scattering
amplitudes and the masses of resonances.  We note that in any such theory,
equation \decomp\ implies that the function $A$ must go to zero at least as
quickly as $1/N$.  Therefore \lowest\ implies that $f$ must go to infinity as
least as fast as $\sqrt{N}$.  If the large-$N$ limit is to be non-trivial,
then $f \to \infty$ {\it exactly} as fast as $\sqrt{N}$, and only $a_0(s)$ is
nonvanishing.  One can turn this around: the ratio of the resonance masses to
$f$ must go like $1/\sqrt{N}$.

Note that this argument would apply to the $SU(N)_L \times SU(N)_R \to
SU(N)_V$ linear $\sigma$-model as well, in contradiction to the claims of
\ref\AT{T. Appelquist and J. Terning, \prd{47}{1993}{3075}.}.  Any model with
a well-defined $N \to \infty$ limit in which there is a low-energy amplitude
enhanced by a factor of $N$ --- no matter what the symmetry group or the
pattern of symmetry breaking --- will have resonances whose masses go like $f
/\sqrt{N}$\ \foot{Of course what one really wants to understand is whether the
masses of resonances in QCD-like theories satisfy a bound like \natural.  For
fixed number of colors, QCD does not have an $N_f \to \infty$ limit, and this
argument does not strictly apply.}.

For simplicity, let us restrict ourselves for the remainder of the paper to
$O(N) \to O(N-1)$ theories in which $A(s,t,u)$ is a function of $s$ only.  We
refer to these as $s$-channel $O(N)$ models.  An example is the $O(N)$ linear
$\sigma$-model.  Such theories can frequently be solved exactly in the limit
of large $N$.  However, that $A(s,t,u)$ depends only on $s$ does not follow
automatically from the large $N$ limit --- it is a very restrictive
condition\foot{It is easy to see that there are theories that have a
well-defined $N \to \infty$ limit that are not of this form.  For example,
consider a scalar field theory with the pions contained in an $O(N)$ vector,
and include an additional symmetric tensor.  When the pions exchange the
tensor field in the $t$ channel the diagram makes a contribution to the
scattering amplitude proportional to $M^2 / (t-M^2)$, a non-trivial function
of $t$.  This model is not soluble even in the limit $N \to \infty$, because
the tensor counts the same way as a gluon in the $N_c \to \infty$ limit of
QCD and all planar diagrams contribute.}.

We will now show that in an $s$-channel $O(N)$ model the most general form of
the function $A$ is \ref\DP{A. Dobado and J. R. Pelaez, \pl{B286}{1992}{136}}
\eqn\DPres{
A(s) = {1 \over {f^2 \over s} F(s;\mu) -
{N \over 32 \pi^2} \log {\mu^2 \over -s}}~~,
}
where $F(s;\mu)$ is an arbitrary function, analytic in a neighborhood of $s=0$
and with
\eqn\fzero{
F(0) = 1~~.
}
Here $\mu$ is the scale at which the theory is renormalized\foot{Ref \DP\
obtained the same form for the scattering amplitude by considering the chiral
Lagrangian for $s$-channel $O(N)$ models.}.

In the limit $N \to \infty$, one has
\eqn\azz{
a_{00}(s) = {N \over 32 \pi} A(s)~~.
}
General principles of field theory tell us that $a_{00}$ is an analytic
function of $s$, everywhere except at values of $s$ corresponding to physical
states.  Below the energy of the lowest massive particle, the only state for
two pions to scatter into is a multipion state.  Since we know that $f^2$ goes
like $N$, we deduce from \Ltwo\ that the vertex for $2 \to 4$ pions is
suppressed by a factor of $1/N^2$ --- and we can easily see that any diagram
for $2 \to 4$ pions is suppressed by exactly the same factor.  There are $N^2$
four-pion states, and therefore the loop computation of the four-pion
contribution to $A$ has two factors of $1/N^2$ at the vertices, but only one
factor of $N^2$ from the intermediate state.  Therefore, the four-pion
contribution to $A$ is a factor of $N$ smaller than the two-pion loop and is
negligible.  Similarly, all other intermediate $2n$ pion states are suppressed
for $n\ge 2$.  From this we see that $a_{00}$ satisfies elastic unitarity in
the $N\to \infty$ limit, and so it lies on the Argand circle for $s$ real,
positive, and less than the lightest massive multi-particle threshold $M_m^2$.
Therefore
\eqn\ampinv{
{\rm Im} {1 \over a_{00}(s)} = -1~~~\hbox{$s$ real, $0<s<M_m^2$}~~,
}
or
\eqn\Ainv{
{\rm Im} {1 \over A(s)} = -{N \over 32 \pi}~~~\hbox{$s$ real, $0<s<M_m^2$}
}
except possibly at a set of isolated points where $a_{00}$ vanishes.

On the other hand, for $s$ real and negative, there is no way to cut an
$s$-channel diagram to yield an on-shell intermediate state.  Thus $A(s)$ and
hence $a_{00}(s)$ must be real.  There should be no poles on the negative real
axis\foot{In the models we consider below, there are tachyonic poles at large
negative real $s$, but they are an indication of the sickness of the theories
at high energy.  In such a case, we focus attention on $|s|$ much less than the
tachyon mass.}.

Now consider the function
\eqn\Gdef{
G(s;\mu) = {1 \over A(s)} +
{N \over 32 \pi^2} \log\left({\mu^2 \over -s}\right)~~.
}

In \fig\cplane{Analyticity of $a_{00}(s)$.  The dots indicate zeros of
$a_{00}$, which correspond to poles in $G$.  The X's indicate the branch
points.  We show that $G$ is analytic in the circle except for simple poles at
the dots.} we show a region of the complex $s$ plane.  The amplitude $a_{00}$
has a branch cut at zero, and possibly also at some point further out the $x$
axis.  These are marked with x's.  Also, $a_{00}$ may have zeros at some
points\foot{We are considering the case where $a_{00}$ has no zeros off the
real axis.  A simple extension of this argument will allow us to handle the
more general case in which $a_{00}$ has zeros off the axis as well.}, which we
indicate by dots in the figure, and this will yield simple poles in $G$.  As
we have seen, for $s$ real and negative, $a_{00}$, and hence $G$, is real.
Therefore, $G(s^*) = G(s)^*$ inside the circle excluding the positive real
axis.  We showed above that the imaginary part of the function $1/A$ went to
$N/(32\pi)$ whenever we approach the positive real axis from above, except at
the points at which $a_{00}(s)$ vanishes.  Therefore the imaginary part of $G$
goes to zero when we approach the positive real axis, whether from above or
below.  Thus, the imaginary part of $G$ is continuous across the $s>0$ axis,
except at the isolated points at which $a_{00}$ vanishes.  By using the
Schwartz reflection theorem\ref\sch{{\it see, for example,} W. Rudin, {\it
Real and Complex Analysis}, McGraw-Hill, New York 1974, p260} on the line
segments between the poles, we see that the function $G$ is analytic
everywhere inside the circle except at the points at which $a_{00}$ vanishes.

We know that there is one point at which $a_{00}$ vanishes and $G$ has a pole,
namely $s=0$.  However, the form of the pole is determined by the low-energy
theorem given in \lowest.  If we now define
\eqn\Fdef{
F(s;\mu) = {s \over f^2} G(s;\mu)~~,
}
then \fzero\ enforces the low-energy theorem\foot{In $O(N)$ models in which
$A$ is not only a function of $s$, the argument given here may be applied to
the isospin zero amplitudes of any angular momentum $\ell$.  In that case, the
function $1/G(s;\mu)$ defined by \Gdef\ has a zero of order at least $\ell$
at $s=0$.  We thank Sidney Coleman for pointing this out.}

\newsec{Convergence of the Effective Theory}

It is of great interest to understand the range of convergence of the
effective theory.  Convergence of the expansion in the effective lagrangian is
intimately related to the existence of poles in the scattering amplitudes.
This discussion will make precise the arguments of ref \us.

To begin, we need to understand the relationship between the form of the
amplitude given in \DPres\ above and the effective Lagrangian describing the
pions.  As mentioned in the previous section, the only diagrams contributing
to $A(s)$ for the pion scattering process at low energy are those that iterate
the two-pion loop in the $s$ channel.  Diagrams with more complicated
topologies, or those that exchange the two-pion loop in the $t$ or $u$
channels are subleading in $N$.  Let us suppose that the tree-level vertex for
$\pi^i \pi^j \to \pi^k \pi^\ell$ scattering in the bare low-energy effective
Lagrangian for an $s$-channel $O(N)$ model is
$iT_0(s) \delta^{ij} \delta^{k\ell} + iT_0(t) \delta^{ik} \delta^{j\ell} +
iT_0(u) \delta^{i\ell} \delta^{jk}$.
To compute pion scattering we simply iterate this vertex with the pion loop.
To do this we need to compute the loop integral
\eqn\Bbardef{
i\overline{B}(p^2;\mu) =
\half \mu^\epsilon \int {d^d\ell\over (2 \pi)^d}
{i \over \ell^2+i \epsilon} {i \over (\ell+p)^2+i \epsilon}
{}~~.
}
This integral is divergent, but the infinity can be absorbed into the tree
level vertex $T_0(s)$, yielding $T(s;\mu)$.  Therefore, we only need to
consider the finite parts of this integral\foot{The renormalization
prescription used here is slightly different from $\overline{MS}$.  The
subtraction is $2/(4-d) - \gamma_E + \log 4\pi - 2$.}:
\eqn\Bdef{
B(s;\mu) =
{1 \over 32\pi^2} \log\left(-{s + i \epsilon \over \mu^2}\right) =
-{1 \over 32\pi^2} \log\left({\mu^2\over -s}\right)
{}~~.
}
In sum, this procedure yields
\eqn\TAres{
A(s) = {T(s;\mu) \over 1 + N T(s;\mu) B(s;\mu)}~~.
}
Therefore, the relationship between the function $F$ defined in the previous
section and the tree-level chiral Lagrangian used here is \DP
\eqn\FTdef{
F(s;\mu) = {s \over f^2 T(s;\mu)}~~.
}

What then is the procedure used in chiral perturbation theory to calculate
pion scattering?  First off, the chiral Lagrangian itself is a power series in
derivatives.  This means that the tree-level vertex is an expansion:
\eqn\Texp{
T(s;\mu) = {s \over f^2}
\sum_{i=0}^\infty t_i(\mu) \left({N s\over f^2}\right)^i~~,
}
where $t_0 = 1$ in order to satisfy the low-energy theorem.  On the other
hand, this expansion is not the only one used in a chiral Lagrangian
calculation.  For example, if one uses the four-derivative vertices in the
chiral Lagrangian (the next terms beyond \Ltwo\ given above), then one must
include the one-loop diagrams as well, since the four-derivative vertices are
one--loop counterterms.  A computation of the amplitude using the chiral
lagrangian including diagrams of up to $\ell$ loops is presented in the form
\eqn\Aldef{
N A_\ell(s) =
\sum_{0<n<m\le \ell+1} c_{mn}(\mu)
\left({N s \over f^2}\right)^m
\left({1 \over 32\pi^2}\log\left({\mu^2 \over -s}\right)\right)^n~~.
}
We recognize this as the expansion of \TAres\ in powers of $s$ and
$\log(\mu/-s)$, truncated in a particular way.  Indeed, we may work out the
coefficients $c_{ij}(\mu)$ by simply expanding the $T$ and $B$ out of the
denominator of \TAres:
\eqn\TAexp{
c_{mn}(\mu) =
\sum_{i_0\ldots i_{m-n-1}} \left({(n+1)! \over i_0! \ldots i_{m-n-1}!}\right)
\prod_{j=0}^{m-n-i} t_j^{i_j}(\mu)
}
where $j$ and $i_j$ are non-negative integers restricted by
\eqn\restr{
\sum_{j=0}^{m-n-1} i_j = n+1 ~~~~\hbox{and}~~~~
\sum_{j=0}^{m-n-1} j i_j = m-n-1 ~~.
}

We see now that there are two distinct but related questions.  First, what is
the radius of convergence of the expansion for $T(s;\mu)$, and second, for
which values of $s$ does
\eqn\Ainfdef{
\lim_{\ell\to \infty} A_\ell(s)
}
exist?

To address the first question we introduce the quantity $m_l$, defined by the
property that $m_l^2$ is the smallest positive\foot{We are once again assuming
that any tachyons are far away from the region of physical interest.} value of
$s$ such that the real part of the amplitude vanishes.  We say that $m_l$ is
the mass of the lightest resonance, since it corresponds to the energy of the
top of the lightest experimentally observed bump in pion scattering.  However,
$m_l^2$ is not the same as the real part of the value of $s$ at which the pole
in $a_{00}$ appears.  Below, we will denote this latter value of $s$ as
$m_p^2 e^{-i \theta_p}$.  If the particle corresponding to $m_l$ is relatively
narrow, then we expect that $m_l$ and $m_p$ are more or less the same.

Let us suppose that $m_l$ is lighter than any other non-analytic structure in
the pion scattering amplitude, such as a multi-particle cut, other than that
from the two-pion loop.  For $s$ smaller than or equal to $m_l^2$, we know
that there are no poles in $T$ off the real axis, since a pole represents a
particle, and, by hypothesis, there are no non-pionic multi-particle states
for the pole to decay into.  Now we choose $\mu = m_l$ in \TAres, and we note
that $B(m^2;m)$ is pure imaginary for any $m$.  Thus, for $A(m_l^2)$ to be
pure imaginary, we see that $T(s;m_l)$ must have a pole at $s=m_l^2$.  This is
in fact the closest pole of $T(s;m_l)$ to $s=0$, and we see that $m_l^2$ is
the radius of convergence of the expansion for $T(s;\mu)$ when $\mu = m_l$.
For other values of $\mu$, we may use the renormalization group to deduce
$t_i(\mu)$ and hence the radius of convergence.  However, since the
coefficients of the chiral Lagrangian depend only logarithmically on $\mu$, we
expect that for reasonable values of $\mu$, neither exponentially large nor
small, the radius of convergence of $T(s;\mu)$ is always about $m_l^2$.

Now we must address the more difficult question of the convergence of the
chiral expansion itself, \ie\ the existence of the limit in \Ainfdef.  We
first note that there is {\it some} region near $s=0$ for it converges.  The
power series for $T(s;m_l)$ is absolutely convergent in the region
$|s|<m_l^2$, furthermore $T(s;m_l)$ goes to zero as fast as $s$ near $s=0$.
Therefore for $s$ in the region $R$ of the complex $s$ plane,
\eqn\abseqn{
R = \left\{s : \sum_{n=0}^\infty
\left|t_n(\mu) \left({N s \over f^2}\right)^{n+1}\right|
\left|{1 \over 32\pi^2}\log\left({\mu^2 \over -s}\right)\right|
< 1\right\}~~,
}
the expansion in \Aldef\ is a sum of terms which would be absolutely
convergent even without the restriction placed by chiral perturbation theory
on the order of their summation \ie\ if we removed the $n<m\le \ell+1$ under
the summation sign.  Therefore, chiral perturbation theory must converge in
$R$.

The chiral perturbation series \Aldef\ is really an expansion of a function
defined on multiple sheets.  Thus, the question of convergence of \Ainfdef\ is
considerably more difficult than that of $T$.  Furthermore, because of the
logarithm terms in the chiral expansion, the usual theorems for convergence of
a series do not apply.  The region $R$ specified in \abseqn\ could be smaller
than the real region of convergence of \Ainfdef.

There are, however, a few educated guesses we can make about the region of
existence of \Ainfdef.  First of all, one sees that when $s=m_l^2$, the chiral
expansion is not {\it absolutely} convergent.  In fact, it seems unlikely that
it can be convergent at all when $|s|>m_l^2$, because the $c_{n0}$ terms by
themselves are growing in magnitude --- $c_{n0}(\mu)$ is just $t_n(\mu)$.
Admittedly the logarithm part of the $c_{nm}$ terms can have an arbitrary
phase, but it seems very unlikely that the radius of convergence of \Ainfdef\
can be substantially larger than $m_l$ on any sheet.

Doubtless the region of convergence of \Ainfdef\ looks something like that
shown in \fig\convfig{The region of convergence of \Ainfdef\ for
complex $s$.  The dot shows the location of the lightest physical pole (on the
second sheet), and the X shows the location of $m_l$.  The dashed line is the
region of convergence on the second sheet.}.  The dot is the location of the
lightest physical pole in pion scattering, at $s=m_p^2 e^{-i\theta_p}$.  This
pole is, as usual, on the second sheet.  The X shows the location of $m_l$,
the pole in tree level scattering.  The form of the region of convergence is
some sort of gentle spiral, furthest from $s=0$ along the negative real axis.
As the logarithm gets a nonzero imaginary part, it grows in magnitude for
fixed $|s|$, and so as we change the argument of $s$ away from the negative
real axis the radius of convergence is reduced.  In the direction of
$\theta_p$, the radius of convergence is exactly equal to $m_p^2$, a number
which is always smaller than or about the same as $m_l^2$.

\newsec{The Natural Effective Theory Criterion in $s$-channel $O(N)$ Models}

Weinberg established a useful definition for the ``naturalness'' of an
effective theory.  As stated above, the chiral Lagrangian is an expansion in
derivatives.  Each term in the expansion serves as a counterterm for the loop
diagrams with vertices from lower order terms.  Weinberg's NET criterion
states that the renormalized coefficients in the higher order chiral
Lagrangian may not be too small.  The argument is as follows.  Consider a
coefficient of a term in ${\cal L}^{2n}$.  In an $s$-channel $O(N)$ model, it
is precisely $t_n(\mu)$.  Since it is a counterterm in a loop diagram,
$t_n(\mu)$ is a running parameter.  Now suppose we change the scale of the
cutoff of the effective theory (or $\mu$ if we are using a dimensional
regulator) by a factor of order 1.  Then $t_n(\mu)$ changes.  If we pick a
particular value of $\mu$, it would be ``unnatural'' to find that $t_n(\mu)$
is very much smaller than, for example, $t_n(e \mu)$ (where $e$ is the base of
the logarithm).  Note that in this usage the word ``natural'' does not have
precisely the same meaning as usual.

The standard naive dimensional analysis \weinnat\ estimate of the minimum size
of $t_k(\mu)$ is $(16 \pi^2)^{-k}$, because a chain of $k$ bubbles has $k$
factors of $\overline B$.  Thus one expects that the radius of convergence of
$T(s;\mu)$, shown in the last section to be $m_l^2$, should be no larger than
about $16 \pi^2 f^2 / N$.  We will see this in explicit examples below.

Interestingly, we may apply the NET criterion directly to the $S$-matrix,
\DPres, and it takes on an especially simple form.  As we saw above, the scale
$\mu$ in the logarithm may be interpreted as the renormalization point of the
infinite pion loop integral.  Therefore, we may ask what happens when we
shift $\mu$ by a factor of $e$.

Since the function $F$ is analytic near $s=0$, we may write
\eqn\Fexpand{
F(s;\mu) = \sum_{k=0}^\infty f_k(\mu) \left({N s \over f^2}\right)^k~~.
}
The factors of $N$ in the numerator above are there because, as we have seen,
$f^2$ counts like a factor of $N$ relative to other dimensionful scales in the
$N \to \infty$ limit.  If there are any terms in the expansion for $F$ that
are not enhanced by a factor of $N^k$, one is not entitled to keep them as
$N \to \infty$.

We note an interesting feature of \Fexpand.  Since $A(s)$ is independent of
$\mu$, we see from \DPres\ that only $f_1$ has any dependence on $\mu$ --- the
others are all constant.  Also, \fzero\ implies that $f_0 = 1$.  It is
important to realize that the reason the $f_k$ may be independent of $\mu$ is
that they are not coefficients of the chiral Lagrangian.  They are derived
from the $t_n(\mu)$'s by a resummation.

We may now apply the NET criterion.  Pick a value of $\mu^2$, some generic
value appropriate for a scattering process.  Since the value of $f_1(\mu)$
would have changed by $1/(16 \pi^2)$ if we had picked $\mu \to e \mu$, we
conclude that $|f_1(\mu)| \gae 1/(16 \pi^2)$.

Therefore, we see that the NET criterion for the effective theory, which is
formulated in terms of the $\beta$ functions of coefficients of the effective
lagrangian, actually puts constraints directly on the $S$-matrix of the full
theory.  We will show below that this implies that new physics, such as a pole
or a branch cut, must enter at a scale below or about $4 \pi f / \sqrt {N}$.

\newsec{Some Examples}

In this section we consider three examples of $s$-channel $O(N)$ models in
order to show how the NET criterion implies that new physics always enters at
a scale at or below $4 \pi f / \sqrt{N}$.

\subsec{The Linear $\sigma$ Model}

We will first consider the $O(N)$ linear $\sigma$ model in the limit of large
$N$, the simplest example of an $s$-channel $O(N)$ model.  This model may be
solved exactly in the $N \to \infty$ limit\ref\CJP{S. Coleman, R. Jackiw, and
H. D. Poltizer, \prd{10}{1974}{2491}.}\ref\Ein{M. B. Einhorn,
\np{B246}{1984}{75}\semi R. Casalbuoni, D. Dominici, and R. Gatto,
\pl{147B}{1984}{419}.}.  The result is that there is a single Higgs resonance
in the ${\rm Re}~s > 0$ region on the second sheet.  Unfortunately, the model
also contains a tachyon, so it cannot be considered an example of an exactly
soluble non-trivial field theory in four dimensions.  We regard the tachyon as
a sign of the fundamental sickness of the theory, and so we must restrict
attention to scales well below the tachyon mass.  If one demands that the mass
of the Higgs resonance is well below the mass of the tachyon, then the Higgs
mass is bounded and one finds that it is less than $4 \pi f / \sqrt{N}$.
Equivalently, if one defines the model with a momentum space cutoff rather
than a dimensional regulator, then there is never a tachyon below the cutoff.
In this case, if we demand that the running coupling constant be finite at the
cutoff, then whenever the Higgs resonance is below the cutoff, it is always
lighter than $4 \pi f / \sqrt{N}$.  For our purposes we regard the tachyon as
being analogous to the momentum cutoff.

In ref \Ein, the relationship between $m_l$ and $m_p$ was considered.  The
tree level vertex is
\eqn\sigtree{
T_0(s) = {s \over f^2} \left({m^2 \over m^2 - s}\right)~~,
}
and therefore
\eqn\flinear{
F(s;\mu) = 1 - \left({f^2 \over N m^2(\mu)}\right)
\left({N s \over f^2}\right) ~~ .
}
where $m(\mu)$ is a parameter with dimensions of mass.  As expected, only the
$f_1$ term depends on $\mu$.  The quantity $m_l$ is the lightest mass such
that the real part of $a_{00}$ vanishes, so as before we choose $\mu = m_l$
and solve for vanishing $F$.  Thus,
\eqn\mreqn{
0 = F(m_l^2;m_l) = 1 - {f^2 \over N m^2(m_l)} {N m_l^2 \over f^2} ~~,
}
so we see that $m^2(m_l) = m_l^2$.  We may find the relationship between $m_p$
and $m_l$ by setting $\mu=m_l$ in \flinear\ and looking for a value of $s =
m_p^2 e^{-i\theta_p}$ such that $A(s)$ has a pole.  One finds the following
simultaneous equations \Ein:
\eqn\Eineqns{
\eqalign{
\cos\theta_p =& \left({m_p^2\over m_l^2}\right)
- {1 \over 2} \left({m_p^2\over m_l^2}\right)
\left({N m_l^2 \over 16 \pi^2 f^2}\right)
\log {m_p^2\over m_l^2} \cr
\sin\theta_p =&
{1 \over 2} \left({m_p^2\over m_l^2}\right) \left({N m_l^2 \over
16 \pi^2 f^2}\right) (\theta_p+\pi)
}
}
Note that there is always a solution at $\theta_p = -\pi$.  This is the
tachyon.  More interesting is ``Higgs remnant'' pole below the real axis on
the second sheet, at positive $\theta_p$.  Whenever $m_t$ is bigger than about
$e m_l$, one finds that $m_l^2 \lae 16 \pi^2 f^2/N$ and $\theta_p \lae \pi/2$.
The ratio $m_p^2/m_l^2$ is always less than 1, ranging from very close to 1
when the theory is weakly coupled ($m_l$ is small), to about .4 when the model
ceases to make sense.

It is instead possible to derive the bound on the resonance mass rather simply
from the NET criterion: the NET criterion applied at $m_l$ is
\eqn\NETone{
|f_1(m_l)| = {f^2 \over N m^2(m_l)} \gae {1 \over 16 \pi^2} ~~,
}
so
\eqn\NETagain{
{16 \pi^2 f^2 \over N} \gae m^2(m_l) = m_l^2 ~~.
}
We see that the NET criterion directly implies that there is an upper bound on
the mass of the resonance.

Derived in this way, the bound on the resonance mass is at least in principle
independent of the existence of the tachyon (or the cutoff).  As we stated
above, this same bound is always obeyed in the $O(N)$ linear $\sigma$ model,
whether or not it satisfies the NET criterion, so long as the tachyon mass,
the cutoff, is less than the mass of the resonance.  This is not too
surprising.  This model is sick at energy scales above the tachyon mass
because the effective potential becomes unbounded below.  In this model we may
understand the naturalness criterion directly as a statement that we stay far
from the sick part of the theory.  Still, it is instructive that the
naturalness of the effective Lagrangian implies something immediately about
the $S$-matrix, without consideration of the effective potential.

\subsec{Arbitrary Numbers of Poles}

One might ask whether it is possible to evade the bound of the previous
example by having more than one pole.  Perhaps the existence of many poles at
a higher mass is adequate to satisfy the NET criterion.  To address this
question we now discuss the $s$-channel $O(N)$ model in which the tree level
expression for $A$ has multiple resonances but no branch cuts, other than the
two-pion cut\foot{We may make such a model by taking including in the
Lagrangian several $O(N)$ singlets, and adding terms that cause them to mix
with the $N^{th}$ component of $\phi$.}.  We will show that any such model
must have at least one resonance whose mass is less than $4 \pi f / \sqrt{N}$.

We take $F^{-1}$ to be a sum of propagators:
\eqn\manyhiggs{
{1 \over F(s;\mu)} = \sum_i a_i(\mu) {m_i^2(\mu) \over m_i^2(\mu) - s} ~~,
}
where the $a_i(\mu)$ are the strengths of the various resonances into $A$.
Since the model has no tachyons or ghosts at tree level, we may assume that at
any reasonable value of $\mu$, for which the theory is well defined, all the
$a_i$ and $m_i^2$ are positive.  Note that \fzero\ implies that $\sum a_i(\mu)
= 1$.

In this case, $F$ is no longer a simple linear function of $s$.  In fact it is
a rational function, and the power series expansion no longer terminates.  We
may compute $f_1$:
\eqn\manybone{
\eqalign{
f_1(\mu)
=&
\left. {f^2 \over N} {d \over ds} F(s;\mu) \right|_{s=0} \cr
=&
\left. -{f^2 \over N} F^2(s;\mu) {d \over ds} F^{-1}(s;\mu) \right|_{s=0} \cr
=&
-\sum_i a_i(\mu) {f^2 \over N m_i^2(\mu)} ~~.
}
}

We now show that the mass of the {\it lightest} resonance must be less than
$4\pi f/\sqrt{N}$.  Again, note that $1/F(s; m_l)$ must have a pole at
$s=m_l$.  From the expression \manyhiggs, we conclude that there must be some
$i$ such that $m_i(m_\ell) = m_\ell$.  Without loss of generality, we may
assume that it is $i=1$.  We now see that $m_1^2(m_\ell)$ is the smallest of
the $m_i^2(m_\ell)$.  Suppose that $m_2^2(m_\ell)$ were lower than
$m_1^2(m_\ell) = m_\ell$.  Start from from $s = m_\ell^2$ and reduce $s$.
Then as we lower $s$ to zero, there must either be a point where
$m_2^2(\sqrt{s}) = s$, or $m_2^2(\sqrt{s}) = -s$.  If the first condition
holds, then we see that the function $F(s;\sqrt{s})$ has a zero at a smaller
value of $s$ than $m_\ell^2$, and thus the amplitude was pure imaginary at a
lower mass than $m_\ell$.  The second condition implies that $F(-s,\sqrt{s})$
has a zero, and so there is a tachyon at a mass scale lower than $m_\ell$.  In
this case we reject the model.

The NET criterion at $\mu = m_l$ states
\eqn\manynat{
{1 \over 16 \pi^2} \lae |f_1(m_l)| =
\sum_i a_i(m_l) {f^2 \over N m_i^2(m_l)} \le
\sum_i a_i(m_l) {f^2 \over N m_1^2(m_l)} = {f^2 \over N m_1^2(m_l)}
{}~~,
}
so
\eqn\manyres{
m_\ell^2 = m_1^2(m_l) \lae {16 \pi^2 f^2 \over N} ~~.
}
Therefore the inclusion of many poles does not permit us to evade the bound on
the mass of the lightest of them.

\subsec{Cuts}

Having concluded that in $s$-channel $O(N)$ models with arbitrary numbers of
poles naturalness implies that there is always a light resonance, we next turn
to a model with a two-particle branch cut.  In this model too there is an
upper bound on the mass of the new physics.

The model we consider has an $O(N) \times O(L)$ symmetry, which breaks to
$O(N-1) \times O(L)$.  The particles $\phi$ transform as a vector under the
$O(N)$ symmetry, and $\psi$ is a vector under $O(L)$.  The Lagrangian is
\eqn\Lcut{
{\cal L} =
  \half \partial^\mu \phi \partial_\mu \phi
+ \half \partial^\mu \psi \partial_\mu \psi
- {\lambda_a\over N} (\phi^2 - v^2)^2
- {\lambda_b\over N} (\psi^2)^2
- {\lambda_c\over N} (\psi^2)(\phi^2 - v^2)
- \half M_\psi^2 \psi^2 ~~.
}
We will solve for the elastic two-to-two particle scattering amplitudes in the
$N \to \infty$ and $L \to \infty$ limit, with $L/N$ held fixed.

So long as $v^2 > 0$ and $M_\psi^2 > 0$, one of the $\phi$'s will get a vacuum
expectation value (VEV), while $\langle \psi \rangle = 0$.  We will take
$\langle \phi_N \rangle = f$, and $\langle \phi_i \rangle = 0$.  The broken
symmetry generators are those $O(N)$ generators $T$ such that $T^{iN} \not =
0$. We will refer to the first $N-1$ $\phi$'s as pions, and the $N^{th}$ as
$\sigma$.

The conditions on the coupling constants such that the tree-level potential
is bounded below are
\eqn\bounded{
\eqalign{
\lambda_a + \lambda_b >& 0 \cr
\lambda_a\lambda_b - \lambda_c^2 >& 0~~, \cr
}
}
from which it follows that both $\lambda_a$ and $\lambda_b$ are positive.  We
will assume that these conditions hold as well for the renormalized couplings
$\lambda(\mu)$, at any renormalization point $\mu$ under consideration.
Unfortunately, this model suffers from the same problem as the linear
$\sigma$ model: at sufficiently large values of $\phi$ the potential becomes
unbounded below.

First consider the diagrams that renormalize the VEV of $\phi_N$.  At tree
level one has the relationship $\langle \phi_N \rangle=v$, but the ``tadpole''
diagrams with an external $\sigma$ will shift the VEV, $f$, relative to the
parameter in the Lagrangian, $v$.  Accordingly, we may eliminate any diagram
with a tadpole sub-diagram, so long as we consistently replace $v$ with $f$.

Next consider the diagrams renormalizing the $\phi$ or $\psi$ propagators.  At
leading $N$ these are ``cactus'' diagrams.  Their effect is to shift the
$\psi$ mass.  There is no $\psi$ or $\pi$ wavefunction renormalization.  We
eliminate any diagram with a cactus subdiagram, and replace $M_\psi$, the mass
parameter in the Lagrangian, with $m_\psi$, the physical $\psi$ mass.

We may now derive the $\pi^i\pi^i\to\pi^j\pi^j$, $i\not= j$ scattering
amplitude as follows.  An arbitrary diagram for this process is a simple chain
of $\sigma$ propagators, $\psi$ loops, and $\pi$ loops (\fig\arbdiag{A
typical diagram contributing to $\pi\pi \to \pi\pi$ scattering.}).  Consider
first the sum of all chains of any number of $\psi$ loops attached together,
with no intervening $\sigma$'s or $\pi$'s.  Having constructed this quantity,
we may sum arbitrary alternations of it with $\sigma$'s.  If we attach such
chains to external $\pi\pi$ states, and include the direct $\pi\pi\pi\pi$
vertex, then we have at this stage all $\pi^i\pi^i \to \pi^j\pi^j$ diagrams in
which the scattering proceeds without internal pion loops.  Therefore, to get
the entire $\pi^i\pi^i \to \pi^j\pi^j$ process, we simply iterate this
scattering with the pion loop.

To do this we need the loop integral for the massive $\psi$'s as well as for
the pions, and thus we introduce the massive loop function\foot{The
subtraction here is the same as for the function $B$ above.}
\eqn\Jdef{
J(s;\mu) =
{1 \over 32\pi^2} \left(
2 +
\int_0^1 dx \log\left({m_\psi^2 - x(1-x)s - i \epsilon\over \mu^2}\right)
\right)
{}~~.
}

With this definition we find that
\eqn\pires{
{1 \over N a_{\pi^i\pi^i\to \pi^j\pi^j}(s)} =
{f^2 \over N s} \left[{1 - {N s \over f^2}
{1 - \lambda_b {L\over N}J(s) \over
\lambda_a + (\lambda_c^2 - \lambda_a \lambda_b){L \over N} J(s)
}}\right] + B(s)
{}~~.
}
(Here we have suppressed the $\mu$ dependence of the quantities $\lambda_a$,
$\lambda_b$, $\lambda_c$, B(s), and $J(s)$.  The VEV $f$ and the function $A$
are physical quantities, independent of $\mu$.)  The quantity in square
brackets may be defined as $F(s;\mu)$, and so we see that this amplitude
satisfies elastic unitarity so long as $J$ is real, which it is for $s <
4m_\psi^2$.

At energies above twice the $\psi$ mass, the $a_{\pi\pi\to\pi\pi}$ no longer
satisfies elastic unitarity, because we may now produce a $\psi\psi$ final
state.  Accordingly, the function $J$, and hence $A$ has a branch cut at $s =
4 m_\psi^2$.  To get $a_{\pi^i\pi^i\to\psi^j\psi^j}$, we use following
procedure.  A diagram for this process is an arbitrary string of pion loops,
$\sigma$ propagators, and $\psi$ loops.  First consider diagrams without $\pi$
loops.  These may be summed by a procedure exactly analogous to that described
above.  Now if the diagram has a pion loop, we divide it in half at its {\it
last} $\pi$ loop.  Such a diagram is therefore the product of a diagram
without a pion loop, a factor of $B$, and an arbitrary $\pi \pi \to \pi \pi$
scattering.  We find in sum:
\eqn\pipsires{
{
a_{\pi^i\pi^i\to\pi^j\pi^j}(s) \over
a_{\pi^i\pi^i \to \psi^j\psi^j}(s)
} =
{\lambda_a\over \lambda_c} +
\left(\lambda_c - {\lambda_a\lambda_b \over \lambda_c} \right) {L\over N} J(s)
{}~~.
}

Lastly, the calculation for $a_{\psi^i\psi^i\to\psi^j\psi^j}$ is precisely
analogous to that of $a_{\pi^i\pi^i\to\pi^j\pi^j}$
\eqn\psires{
{1 \over N a_{\psi^i\psi^i\to \psi^j\psi^j}(s)} =
{
\lambda_a\left({f^2 \over N} + s B(s)\right) - s
\over
\lambda_b s - (\lambda_a \lambda_b - \lambda_c^2)
\left({f^2 \over N} + s B(s)\right)
}
+ {L\over N} J(s)
{}~~.
}

We may now derive three renormalization group equations for the three coupling
constants.  First, note that since all amplitudes are independent of $\mu$,
the quantity on the right hand side of \pipsires\ must be independent of $\mu$.
Since
\eqn\BJderiv{
{\partial J(s;\mu)\over\partial\log\mu} =
{\partial B(s;\mu)\over\partial\log\mu} =
-{1\over 16 \pi^2}
}
is independent of $s$, we conclude that
\eqn\RGa{
{\partial \over \partial\log\mu}
\left(
{\lambda_a\lambda_b \over \lambda_c} - \lambda_c
\right)
= 0
{}~~.
}
We refer to this particular invariant combination of coupling constants as
$\gamma$.  The invariance of the entire right hand side of \pipsires\ implies
that
\eqn\RGb{
{\partial \over \partial\log\mu} {\lambda_a\over \lambda_c}
- \gamma {L\over N} {\partial \over \partial\log\mu} J(s)
= 0
{}~~.
}
Lastly, the invariance of the $\pi\pi\to\pi\pi$ scattering amplitude in
\pires\ implies
\eqn\RGc{
{\partial \over \partial\log\mu} {\lambda_b\over \lambda_c}
- \gamma {\partial \over \partial\log\mu} B(s)
= 0
{}~~.
}
We may rewrite these equations as
\eqn\RG{
\eqalign{
{\partial \over \partial\log\mu} \left({1\over \lambda_a}\right) &=
-{1 \over 16\pi^2}
\left(1 + {L\over N} {\lambda_c^2\over \lambda_a^2}\right)
\cr
{\partial \over \partial\log\mu} \left({1\over \lambda_b}\right) &=
-{1 \over 16\pi^2}
\left({L\over N} + {\lambda_c^2\over \lambda_b^2}\right)
\cr
{\partial \over \partial\log\mu} \left({1\over \lambda_c}\right) &=
-{1 \over 16\pi^2}
\left({L\over N} {\lambda_b\over \lambda_c} + {\lambda_a\over
\lambda_c}\right)
{}~~.
}
}
The solutions to these equations are
\eqn\RGsol{
\eqalign{
{\lambda_a \over \lambda_c} =&
{\gamma\over 16\pi^2}{L \over N} \log{\Lambda_a \over \mu} \cr
{\lambda_b \over \lambda_c} =&
{\gamma\over 16\pi^2} \log{\Lambda_b \over \mu} \cr
{1 \over \lambda_c} =&
{\gamma \over (16 \pi^2)^2}{L \over N} \log{\Lambda_a \over \mu}
\log{\Lambda_b \over \mu} - {1 \over \gamma}
{}~~.
}
}
Therefore, this model can be parametrized by two dimensionful quantities,
$\Lambda_a$ and $\Lambda_b$, and one dimensionless combination of couplings,
$\gamma$.

The content of the NET criterion is
\eqn\bonecut{
{1 \over 16\pi^2} \le |f_1(\mu)| =
{1 - \lambda_b(\mu) {L\over N}J(0;\mu) \over
\lambda_a(\mu) + (\lambda_c(\mu)^2 - \lambda_a(\mu) \lambda_b(\mu))
{L \over N} J(0;\mu)
}
}
The content the NET criterion is therefore that this equation must hold for
any scale $\mu$ in which the theory is used to calculate scattering.  In
particular, we expect that it holds both at $\mu = m_l$ and at $\mu = m_\psi$.
We may plug in the solutions to the renormalization group equations and the
definition of $J$ to find
\eqn\NETcut{
\Gamma \le
\left(\log {\Lambda_a \over m_\psi} - 1 \right)
\left(\log {\Lambda_b \over \mu} - 1 \right)
}
where $\Gamma = ((16 \pi^2)^2 N)/(\gamma^2 L)$.  For this to hold
we must either have $\log{\Lambda_a/m_\psi}$ and
$\log{\Lambda_b/\mu}$ both greater than 1 or both less than zero.  The latter
possibility is excluded, because then $\lambda_a$ and $\lambda_b$ themselves
are negative, and the potential is unbounded below at a small scale, $\mu$ or
$m_\psi$.

Now we will solve for $m_l$, using $F(m_l^2;m_l) = 0$.  Using the form of $F$
given in \pires, we find that the equation for $m_l$ is
\eqn\mreqn{
{N m_l^2 \over f^2} = \lambda_a(m_l) +
{\lambda_c^2(m_l) J(m_l^2;m_l) \over 1 - \lambda_b(m_l) J(m_l^2;m_l)}
{}~~.
}
We now want to know what the natural scale for $m_l$ is.

We now plug the solutions of the renormalization group equations into \mreqn:
\eqn\mreqntwo{
{N m_l^2 \over f^2} =
16 \pi^2
{
\displaystyle \log{\Lambda_a\over m_l} - 16 \pi^2 J(m_l^2;m_l)
\over
\displaystyle \log{\Lambda_a\over m_l} \log{\Lambda_b\over m_l}
- \Gamma - \log{\Lambda_b\over m_l} 16 \pi^2 J(m_l^2;m_l)
}
{}~~.
}
We write this as
\eqn\mreqnthree{
{m_l^2 \over \left({16 \pi^2 f^2 \over N}\right)}
=
\left(
\log{\Lambda_b\over m_l} -
{\displaystyle\Gamma
 \over
 \displaystyle \log{\Lambda_a\over m_l} - 16 \pi^2 J(m_l^2;m_l)
}
\right)^{-1}
{}~~.
}
We now show that this equation has a root in the range $0 < m_l^2 < 16 \pi^2
f^2 / N$.  The left hand side grows linearly with $m_l^2$ and reaches one when
$m_l^2 = 16 \pi^2 f^2 / N$.  We will show below that the right hand side has
infinite derivative at $m_l^2 = 0$, and it is bounded by one for $m_l < 2
m_\psi$.  Therefore, when $m_l^2$ is just above zero the right hand side is
bigger than the left hand side, but it must fall below the left hand side
before $m_l^2 = 16 \pi^2 f^2 / N$, so long as $(2 m_\psi)^2 > 16\pi^2 f^2 / N$.

All that remains is to show the two assertions of the previous paragraph.
First, we note that we may write
\eqn\Jeqn{
16 \pi^2 J(m_l^2, m_l) = \log\left({m_\psi \over m_l}\right) +
g\left({m_l^2\over 4 m_\psi^2}\right)~~.
}
Here $g(x)$ is a real analytic function for $x<1$.  The function $g$ is
monotonically decreasing as $x$ increases: $g(0) = 1$ and $g(1) = 0$.
Now the right hand side of \mreqnthree\ is
\eqn\Reqn{
\left(\log{\Lambda_b\over m_l} -
{\displaystyle\Gamma
 \over
 \displaystyle \log{\Lambda_a\over m_l} - 16 \pi^2 J(m_l^2;m_l)
}
\right)^{-1}
=
\left(
\log{\Lambda_b\over m_l} -
{\displaystyle\Gamma
 \over
 \displaystyle \log{\Lambda_a\over m_\psi}
 - g\left({m_l^2\over 4 m_\psi^2}\right)
}
\right)^{-1}
{}~~.
}
Since $g$ is analytic as $m_l\to 0$, the whole expression is well approximated
by $(\log \Lambda_b/m_l)^{-1}$, and the derivative blows up.

To show the second assertion, we note that
\eqn\boundeqn{
\left(
\log{\Lambda_b\over m_l} -
{\displaystyle\Gamma
 \over
 \displaystyle \log{\Lambda_a\over m_\psi}
 - g\left({m_l^2\over 4 m_\psi^2}\right)
}
\right)^{-1}
\le
\left(
\log{\Lambda_b\over m_l} -
{\displaystyle\Gamma
 \over
 \displaystyle \log{\Lambda_a\over m_\psi} - 1
}
\right)^{-1}
{}~~.
}
so long as $m_l < 2 m_\psi$.  The right hand side is less than one if
\eqn\boundcondit{
\left(\log{\Lambda_a\over m_\psi} - 1 \right)
\left(\log{\Lambda_a\over m_l} - 1 \right)
> \Gamma
{}~~.
}
This must hold, because it is just the NET criterion with $\mu = m_l$.

In this section we have considered three explicit examples of $s$-channel
$O(N)$ models.  A general $s$-channel $O(N)$ model is never too much more
complicated - one can simultaneously add multiple poles and branch cuts.  In
the models of this section either there is a resonance at a mass $m_l$ below
$4 \pi f / \sqrt{N}$, or there is a branch cut before that scale.  Some new
physics must always enter --- the addition of multiple poles or of a branch cut
did not permit us to evade the bound.

\newsec{Conclusion}

In this paper we have considered a type of strongly interacting field theory.
We have seen that the tree-level chiral lagrangian for the pions converges
only out to the mass of the lightest resonance, and that the chiral loop
computations converge to a scale of this order or smaller.

In addition, we have shown that the natural effective theory criterion implies
something about the $S$-matrix of the full theory.  In the cases considered
here, we have shown that there must always be new physics, a resonance or a
cut, at or below $4 \pi f / \sqrt{N}$.  This bound holds independent of the
precise form of the full theory or of the detailed consideration of the
effective potential.

The theories considered here are really only toy models.  Because of their
simplicity, it was directly possible to interpret the natural effective theory
criterion as the condition that keeps us far from the sick part of the theory,
as indicated either by an explicit momentum space cutoff or the tachyon.
Accordingly, in the cases considered here, it was not possible to construct
{\it any} theory with an unnatural effective theory.  The open question is,
therefore, whether the NET criterion is a general property of field theory, or
if there are models that somehow evade it and have resonances that are heavy
compared to the naive scale.

\bigskip

\noindent
{\bf Acknowledgments}

\medskip

We thank R. Sekhar Chivukula, Sidney Coleman, Howard Georgi, Thorsten Ohl, and
John Terning for useful discussions.  M.G. thanks the Texas National Research
Laboratory Commission for support under a Superconducting Super Collider
National Fellowship. This research is supported in part by the National
Science Foundation under grant PHY--92--18167, by the Texas National Research
Laboratory Commission, under grants RGFY93--278 and RGFY93--278B, and by the
Department of Energy, under contract number DE--FG02--91ER40676.

\listrefs
\listfigs

\message{Now producing pictures.  Warning: pictex is slow!}
\input pictex

$$
\beginpicture
\setlinear
\setcoordinatesystem units <1cm,1cm>
\plot 4.5 5 4.5 4.5 5 4.5 /
\put {$s$} at 4.75 4.75
\plot -5 0 5 0 /
\plot 0 -5 0 5 /
\plot -.1 .1 .1 -.1 /
\plot -.1 -.1 .1 .1 /
\plot 3.9 .1 4.1 -.1 /
\plot 3.9 -.1 4.1 .1 /
\put {$\bullet$} at 0 0
\put {$\bullet$} at 1 0
\put {$\bullet$} at 3 0
\setdashes
\plot 0 -.08 5 -.08 /
\plot 4 -.04 5 -.04 /
\setsolid
\circulararc 358 degrees from 4 0 center at 0 0
\linethickness=0pt
\putrule from -5 0 to 5 0
\putrule from 0 -5 to 0 5
\endpicture
$$
\vfil\eject
$$
\beginpicture
\setlinear
\setcoordinatesystem units <1cm,1cm>
\plot 4.5 5 4.5 4.5 5 4.5 /
\put {$s$} at 4.75 4.75
\plot -5 0 5 0 /
\plot 0 -5 0 5 /
\plot 3.9 .1 4.1 -.1 /
\plot 3.9 -.1 4.1 .1 /
\plot
 3.45 0
 3.46 -0.14
 3.47 -0.32
 3.46 -0.49
 3.45 -0.67
 3.43 -0.85
 3.40 -1.02
 3.36 -1.19
 3.31 -1.37
 3.25 -1.54
 3.18 -1.70
 3.11 -1.87
 3.02 -2.03
 2.93 -2.19
 2.82 -2.34
 2.71 -2.49
 2.60 -2.63
 2.47 -2.76
 2.34 -2.89
 2.20 -3.02
 2.05 -3.13
 1.90 -3.24
 1.74 -3.34
 1.57 -3.44
 1.40 -3.52
 1.23 -3.60
 1.05 -3.66
 0.87 -3.72
 0.68 -3.77
 0.49 -3.81
 0.30 -3.84
 0.11 -3.86
-0.08 -3.87
-0.27 -3.87
-0.47 -3.85
-0.66 -3.83
-0.85 -3.80
-1.04 -3.76
-1.23 -3.71
-1.42 -3.65
-1.60 -3.58
-1.78 -3.50
-1.96 -3.42
-2.13 -3.32
-2.30 -3.21
-2.46 -3.10
-2.61 -2.97
-2.76 -2.84
-2.90 -2.71
-3.04 -2.56
-3.17 -2.41
-3.28 -2.25
-3.40 -2.08
-3.50 -1.91
-3.59 -1.73
-3.68 -1.55
-3.75 -1.37
-3.82 -1.18
-3.87 -0.99
-3.92 -0.79
-3.95 -0.60
-3.98 -0.40
-3.99 -0.20
-4.00  0.00
-3.99  0.20
-3.98  0.40
-3.95  0.60
-3.92  0.79
-3.87  0.99
-3.82  1.18
-3.75  1.37
-3.68  1.55
-3.59  1.73
-3.50  1.91
-3.40  2.08
-3.28  2.25
-3.17  2.41
-3.04  2.56
-2.90  2.71
-2.76  2.84
-2.61  2.97
-2.46  3.10
-2.30  3.21
-2.13  3.32
-1.96  3.42
-1.78  3.50
-1.60  3.58
-1.42  3.65
-1.23  3.71
-1.04  3.76
-0.85  3.80
-0.66  3.83
-0.47  3.85
-0.27  3.87
-0.08  3.87
 0.11  3.86
 0.30  3.84
 0.49  3.81
 0.68  3.77
 0.87  3.72
 1.05  3.66
 1.23  3.60
 1.40  3.52
 1.57  3.44
 1.74  3.34
 1.90  3.24
 2.05  3.13
 2.20  3.02
 2.34  2.89
 2.47  2.76
 2.60  2.63
 2.71  2.49
 2.82  2.34
 2.93  2.19
 3.02  2.03
 3.11  1.87
 3.18  1.70
 3.25  1.54
 3.31  1.37
 3.36  1.19
 3.40  1.02
 3.43  0.85
 3.45  0.67
 3.46  0.49
 3.47  0.32
 3.46  0.14
 3.45  0
/
\setdashes
\plot
 3.45 0
 3.43 -0.20
 3.39 -0.37
 3.35 -0.54
 3.30 -0.70
 3.25 -0.86
 3.18 -1.01
 3.11 -1.16
 3.03 -1.31
 2.94 -1.45
 2.85 -1.59
 2.75 -1.72
 2.64 -1.84
 2.53 -1.96
 2.41 -2.07
 2.29 -2.17
 2.16 -2.27
 2.03 -2.35
/
\put{$\bullet$} at  3.18 -1.01
\linethickness=0pt
\putrule from -5 0 to 5 0
\putrule from 0 -5 to 0 5
\endpicture
$$
\vfil\eject
$$
\beginpicture
\setcoordinatesystem units <1cm,1cm>
\setlinear
\plot 0 0 1 1 0 2 /
\circulararc 360 degrees from 2 1 center at 1.5 1
\circulararc 360 degrees from 3 1 center at 2.5 1
\plot 3 1 4 1 /
\circulararc 360 degrees from 5 1 center at 4.5 1
\put {$\ldots$} at 5.5 1
\circulararc 360 degrees from 7 1 center at 6.5 1
\plot 7 1 8 1 /
\plot 9 0 8 1 9 2 /
\put {$\pi^i$} at -.2 0
\put {$\pi^j$} at -.2 2
\put {$\pi^k$} at 9.2 0
\put {$\pi^l$} at 9.2 2
\put {$\psi$} at 1.5 1.7
\put {$\psi$} at 1.5 0.3
\put {$\pi$} at 2.5 1.7
\put {$\pi$} at 2.5 0.3
\put {$\sigma$} at 3.5 1.2
\put {$\psi$} at 4.5 1.7
\put {$\psi$} at 4.5 0.3
\put {$\pi$} at 6.5 1.7
\put {$\pi$} at 6.5 0.3
\put {$\sigma$} at 7.5 1.2
\linethickness=0pt
\putrule from 0 0 to 9 0
\putrule from 0 0 to 0 2
\endpicture
$$

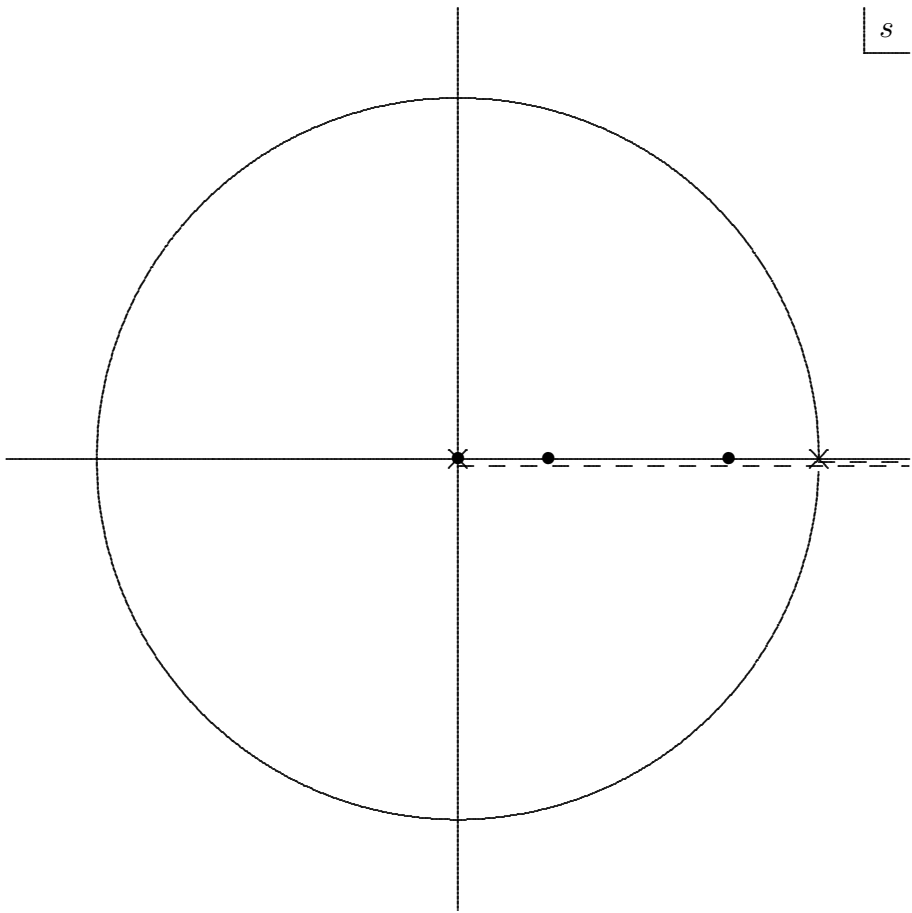
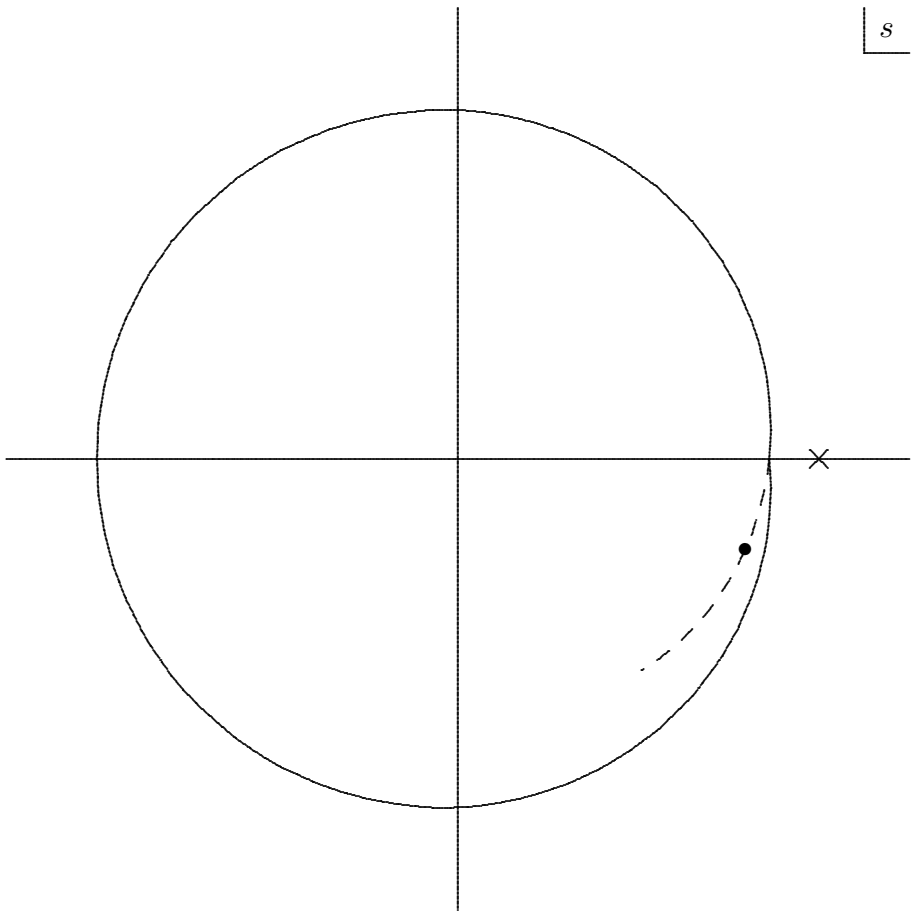
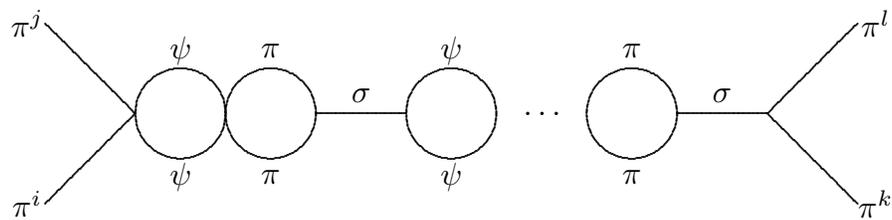
\bye